\documentclass[12pt, epsf]{JHEP3}

\usepackage{epsfig}

\def\pic #1#2{\hbox{\lower#1pt\hbox{\mbox{\epsfxsize=16truemm 
\epsffile{#2}}}}}
\def\picl #1#2{\hbox{\lower#1pt\hbox{\mbox{\epsfxsize=30truemm 
\epsffile{#2}}}}}

\def\picel #1#2{\hbox{\lower#1pt\hbox{\mbox{\epsfxsize=50truemm 
\epsffile{#2}}}}}

\newcommand{\non}{\nonumber \\*}

\newcommand{\eq}[1]{eq.~(\ref{#1})}

\def\bea{\begin{eqnarray}}
\def\eea{\end{eqnarray}}

\def\LB{\left(}
\def\RB{\right)}
\def\be{\begin{equation}}
\def\ee{\end{equation}}

\def\ex{{\rm exp}}

\def\C{{\cal C}}

\def \bi{\bibitem}

\def\au{{\rm Aut}(G)}
\def\Z{{\cal Z}}

\title{\LARGE Non-Abelian Wilson Surfaces}
\author{Iouri Chepelev \\Department of Physics, California Institute of Technology\\Pasadena, CA 91125\\and\\
C.N. Yang Institute for Theoretical Physics\\
SUNY at Stony Brook, NY 11794\\Email: \email{chepelev@theory.caltech.edu}}

\abstract{A definition of non-abelian genus zero open 
Wilson surfaces is proposed.
The ambiguity in surface-ordering is compensated by the gauge
transformations.}
\preprint{CALT-68-2357\\ YITP-SB-01-66}

\begin{document}

\section{Introduction}
A higher dimensional generalization of the non-abelian Wilson line
is not known. 
Only recently the notion of a connection on a
non-abelian 1-gerbe was introduced  in the
work of Breen and Messing \cite{BM}.

A motivation for  defining the  Non-abelian Wilson
Surfaces   comes  from  the string theory.  
NWS  are relevant to
six dimensional  theories
on the world volumes of coincident five branes \cite{six}.

The main problem in defining  NWS is the lack of a  natural
order on a 2-dimensional surface.
A naive guess for the NWS is
\be
P ~\ex \LB \int_{\Sigma} B\RB , 
\ee
where $B$ is a non-abelian 2-form.
The choice of  a surface-ordering $P$ involves a time-slicing of the 2-surface
$\Sigma$.
A no-go theorem of Teitelboim \cite{teitel} states that no such a choice is
compatible with  the 
reparametrization invariance.

Let us recall the notion of a connection on a non-abelian
 1-gerbe \cite{BM}.
A connection on a principal bundle (0-gerbe) can be 
thought of as follows. Let $x_0$ and $x_1$ be two infinitesimally close
points. The fibers $S_{x_0}$ and $S_{x_1}$ over these points are sets
and the connection is  a function
\be
f_{01}: S_{x_1} \rightarrow S_{x_0}.
\ee
The connection on a non-abelian
 1-gerbe is defined by analogy with the 0-gerbe case 
\cite{BM}.
The fibers are  categories $C_{x_0}$ and $C_{x_1}$, 
and the connection is a functor
\be
\varepsilon_{01}: C_{x_1} \rightarrow C_{x_0}.
\ee
Let $x_0$, $x_1$ and $x_2$ be three infinitesimally close points.
A diagram of functors and natural transformations is shown in figure 
\ref{natural}.
Let $\au$ be the group of automorphisms of a non-abelian group $G$. Let
Lie$(G)$ be the Lie algebra of $G$. 
It is shown in \cite{BM} that   2-arrow $K$,  1-arrow $\kappa$ and
1-arrow $\varepsilon$ in the
diagram correspond to a Lie($G$)-valued 2-form $B$, a Lie($\au$)-valued
 2-form $\nu$ and a Lie($\au$)-valued 1-form $\mu$ respectively.

The paper is organized as follows. 
In section 2 a definition of NWS is proposed. Section 3 is devoted to
gauge transformations. Some comments are listed in section 4.

\FIGURE[ht]{
\includegraphics[width=4truecm]{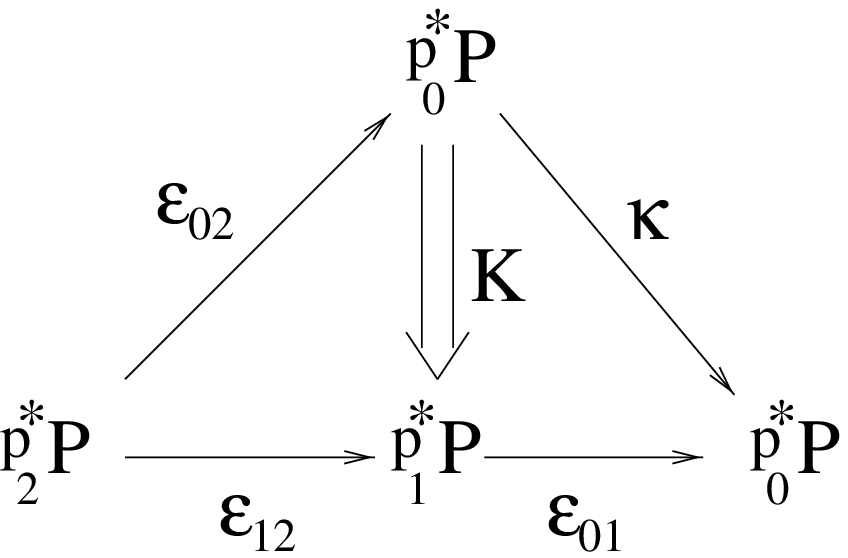}
\hspace{.5truecm}

\centerline{}
{\vspace{-8truemm}}
\caption{$\varepsilon_{ij}$ is a Cartesian functor from the fibered category 
$p_j^* P$ to $p_i^*P$, $\kappa$ is a Cartesian functor from 
$p_0^* P$ to $p_0^* P$,
and $K$ is a 2-arrow from $\kappa \circ\varepsilon_{02}$ to 
$\varepsilon_{01} \circ\varepsilon_{12}$.}
\label{natural}
}

\section{Definition}
We interpret the infinitesimal 2-simplex in  figure \ref{natural} as
a transmuted form of
an infinitesimal Wilson surface expressed in the language of category
theory. The fibered category in the formulation of \cite{BM} can
be thought of as an `internal symmetry space' of a non-abelian string. 
Let $\Sigma$ be a 2-dimensional surface with the disk topology. Let $C$
be a clockwise oriented boundary of $\Sigma$ and $P$ a marked point on 
it (see figure \ref{compose}).
We associate  group elements 
$$W[\Sigma, C, P]\in G$$
and
$$V[\Sigma, C, P]\in \au$$
with the data $(\Sigma, C, P)$. We   write 
$W[\Sigma]$ and $V[\Sigma]$ when the omitted arguments are obvious
from the context. 
With an open curve $C$ we associate an element of $\au$:
\be
M[C] \in \au .
\ee
Let $C=C_2\circ C_1$ be a composition of curves $C_2$ and $C_1$. 
We assume
\be
M[C]= M[C_2 \circ C_1]= M[C_2] M[C_1] .
\ee

We now propose an equation relating $M[C]$, $W[\Sigma,C]$
and $V[\Sigma,C]$.  
For a group element $g\in G$ we denote by $i_g$ the inner automorphism
\be
i_g(h) = g h g^{-1},~~~~\forall h\in G .
\ee
The conjectural equation reads
\be
M[C] = i_{W[\Sigma]} V[\Sigma] .
\label{fund}
\ee
An infinitesimal version of this equation was first derived in \cite{BM}
from the requirement that $K$ in figure \ref{natural} is 
a natural transformation. We regard \eq{fund} as a fundamental equation
relating bulk and boundary  of the non-abelian string 
world-sheet. 

Eq.(\ref{fund}) can be used to find a composition rule for two NWS.
Consider  the 2-surfaces  in figure \ref{compose}. The identity
\bea
i_{W[\Sigma_2\circ \Sigma_1,P_1]}V[\Sigma_2\circ \Sigma_1]&=& M[C\circ C_4 \circ C_3]\non
&=& M[C] M[C_4\circ C_5^{-1}] M[C^{-1}] M[C\circ C_5 
\circ C_3] \non
&=& M[C] i_{W[\Sigma_2, P_2]} V[\Sigma_2,P_2] M[C^{-1}] i_{W[\Sigma_1,P_1]}
V[\Sigma_1,P_1]
\eea
suggests the following
composition rule for Wilson surfaces:
$$
W[\Sigma_2\circ \Sigma_1] = M[C]( W[\Sigma_2] ) 
M[C] V[\Sigma_2] M[C^{-1}] ( W[\Sigma_1] ),
$$
\be
V[\Sigma_2\circ \Sigma_1] = M[C] V[\Sigma_2] M[C^{-1}] V[\Sigma_1].
\label{funcom}
\ee
An infinitesimal version of \eq{funcom} appeared implicitly
in the category-theoretic definition of the curvature in \cite{BM}.

\FIGURE[ht]{
\includegraphics[width=12truecm]{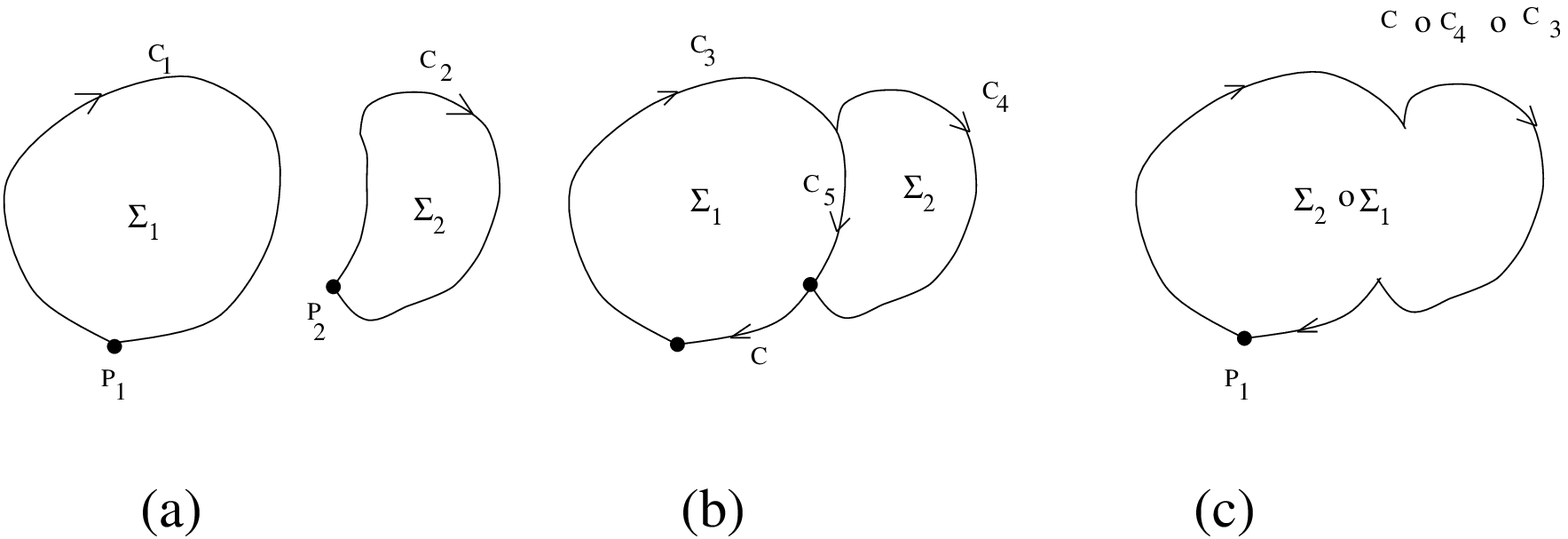}
\hspace{.5truecm}

\centerline{}
{\vspace{-8truemm}}
\caption{Composition of surfaces with the disk topology. (a) Surfaces 
$\Sigma_i$ with the marked points $P_i$ and the clockwise oriented 
boundaries $C_i$. (b) Surfaces are joined along the common boundary segment
$C_5$. (c) The resulting surface $\Sigma_2 \circ \Sigma_1$ with the marked
point $P_1$ and the clockwise oriented boundary $C\circ C_4\circ C_3$.}
\label{compose}
}

Eq.(\ref{funcom}) can be understood as follows.
When the curve $C$ is absent, i.e. when the marked points of $\Sigma_1$ and
$\Sigma_2$ coincide, \eq{funcom} simplifies to
$$
W[\Sigma_2\circ \Sigma_1] =  W[\Sigma_2]  
V[\Sigma_2]( W[\Sigma_1] ),
$$
\be
V[\Sigma_2\circ \Sigma_1]= V[\Sigma_2]  V[\Sigma_1].
\label{funcom1}
\ee
Thus when the marked points of the two surfaces coincide, the Wilson
surfaces are composed as in \eq{funcom1}. If we think of $V[\Sigma,P]$ 
as an operator which acts on the objects with the marked point $P$ 
and  assume that only the objects with the same marked points can
be multiplied, then the meaning of \eq{funcom} becomes clear. The role
of $M[C]$ in \eq{funcom} is to transform the objects with the marked point 
$P_2$ to the objects with the marked point $P_1$.

\FIGURE[ht]{
\includegraphics[width=1truecm]{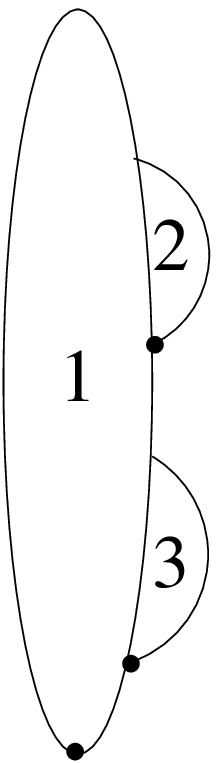}
\hspace{.5truecm}

\centerline{}
{\vspace{-8truemm}}
\caption{$\Sigma_3 \circ (\Sigma_2\circ \Sigma_1) \ne \Sigma_2 \circ 
(\Sigma_3 \circ \Sigma_1)$
}
\label{notcom}
}

Composition of  three or more
surfaces  is in general  ambiguous.
 Consider figure \ref{notcom}. Using the composition rule
(\ref{funcom}) it can  be shown that 
$$
W[\Sigma_3 \circ (\Sigma_2\circ \Sigma_1)] \ne W[\Sigma_2 \circ 
(\Sigma_3 \circ \Sigma_1)],
$$
\be
V[\Sigma_3 \circ (\Sigma_2\circ \Sigma_1)] \ne V[\Sigma_2 \circ 
(\Sigma_3 \circ \Sigma_1)].
\ee

Given 
\be
V[\delta \Sigma]\approx 1+ v[P] \equiv 1 + v_{\mu\nu}[P] \sigma^{\mu\nu}
\ee
 for an infinitesimal surface $\delta \Sigma$ with the area element
$\sigma^{\mu\nu}$,
we want to find $V[\Sigma]$ for a finite-size surface $\Sigma$. 
This can be done using a trick similar to the one used in the context of
the non-abelian Stokes formula \cite{stokes}.  
Consider the contour $C'$ in figure \ref{nonabelian}. From the relation
\be
M[C']= M[C_P^{-1}] M[\delta C] M[C_P] M[C]
\ee
and \eq{fund} one finds
\be
V[\Sigma']= M[C_P^{-1}] V^{-1}[\delta \Sigma] M[C_P] V[\Sigma].
\ee
Thus we have
\be
\delta V[\Sigma] = M[C_P^{-1}] v[P] M[C_P] V[\Sigma].
\ee 
A solution of this equation involves a choice of ordering and
it is given by 
\be
V[\Sigma] = {\hat P}_{\tau} \exp \LB \int_{\Sigma} M[C_P^{-1}] v[P] M[C_P]
 \RB ,
\label{Vsurface}
\ee
where ${\hat P}_{\tau}$ is the ordering in $\tau$ and the curve $C_P$ is
defined in figure \ref{ordering}. Note that the expression \eq{Vsurface}
depends on the parametrization $x^{\mu}= x^{\mu}(\sigma, \tau)$ of the 
surface $\Sigma$. For example a boundary-preserving reparametrization 
will change  $C_P$ to a $C'_P$ (see figure \ref{ordering}). 
Thus $V[\Sigma]$ and $W[\Sigma]$ depend on the parametrization of
$\Sigma$: 
\be
V=V[\Sigma,x^{\mu}(\sigma,\tau)],~~~W=W[\Sigma, x^{\mu}(\sigma,\tau)].
\ee
In  section 3 we will see that if $(\sigma,\tau)$ and $({\tilde \sigma},{\tilde \tau})$ 
are two different parametrizations of a  surface $\Sigma$, 
then
$$(V[\Sigma,x^{\mu}(\sigma,\tau)],W[\Sigma,x^{\mu}(\sigma,\tau)])$$
 and 
$$(V[\Sigma,x^{\mu}({\tilde \sigma},{\tilde \tau})],W[\Sigma,x^{\mu}({\tilde \sigma},{\tilde \tau})])$$
 are related by the gauge transformation. 
In other words, the non-abelian internal symmetry and the
 reparametrization symmetry mix.

\FIGURE[ht]{
\includegraphics[width=5truecm]{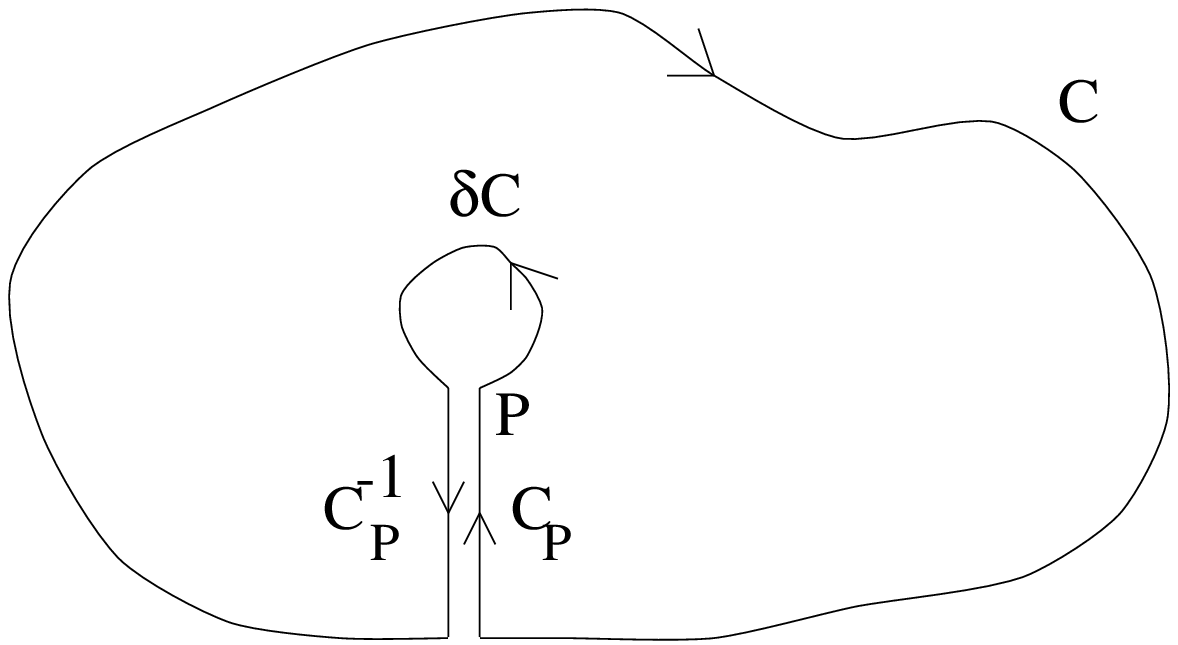}
\hspace{.5truecm}

\centerline{}
{\vspace{-8truemm}}
\caption{Contour $C'= C_P^{-1} \circ \delta C \circ C_P \circ C$.}
\label{nonabelian}
}

\section{Gauge transformations} 
In this section 
we  introduce the gauge transformations which  compensate
 the ambiguity in the composition of NWS. 
Suppose that a surface $\Sigma$ is composed out of three or more 
smaller surfaces. 
Let $(W[\Sigma],V[\Sigma])$ and  $({\tilde W}[\Sigma],{\tilde V}
[\Sigma])$ correspond  to two different compositions resulting in the 
surface $\Sigma$. We have
\be
M[C] = i_{W[\Sigma]} V[\Sigma] = i_{{\tilde W}[\Sigma]} {\tilde V}[\Sigma].
\label{fun1}
\ee

Since $W$ and ${\tilde W}$ are elements of a group $G$, 
there is a group element $R[\Sigma]\in G$ such that
\be
{\tilde W}[\Sigma] = W[\Sigma] (R[\Sigma])^{-1}.
\label{gauge}
\ee
Let us decompose $W$ and ${\tilde W}$ into the abelian and non-abelian
factors:
\be
W = W_{\scriptstyle{\rm ab}} \cdot W_{\scriptstyle{\rm nonab}},~~~~~
{\tilde W}= 
{\tilde W}_{\scriptstyle{\rm ab}} \cdot {\tilde W}_{\scriptstyle{\rm nonab}}.
\ee
It is clear that  the ambiguity in the composition does not affect
the abelian part. Thus we have
\be
{\tilde W}_{\scriptstyle{\rm ab}}[\Sigma]=W_{\scriptstyle{\rm ab}} [\Sigma].
\label{gauge1}
\ee
Combining this equation with \eq{gauge} we find
\be
{\tilde W}_{\scriptstyle{\rm nonab}}[\Sigma] = W_{\scriptstyle{\rm nonab}}
[\Sigma] (R[\Sigma])^{-1}.
\label{gauge2}
\ee
We propose that \eq{gauge1} and \eq{gauge2} define the gauge transformation
of $W$. In order for this gauge transformation of $W$ to be compatible with
\eq{fun1},  $V$  should transform as 
\be
{\tilde V}[\Sigma] = i_{R[\Sigma]} V[\Sigma].
\label{gauge3}
\ee

\FIGURE[ht]{
\includegraphics[width=6truecm]{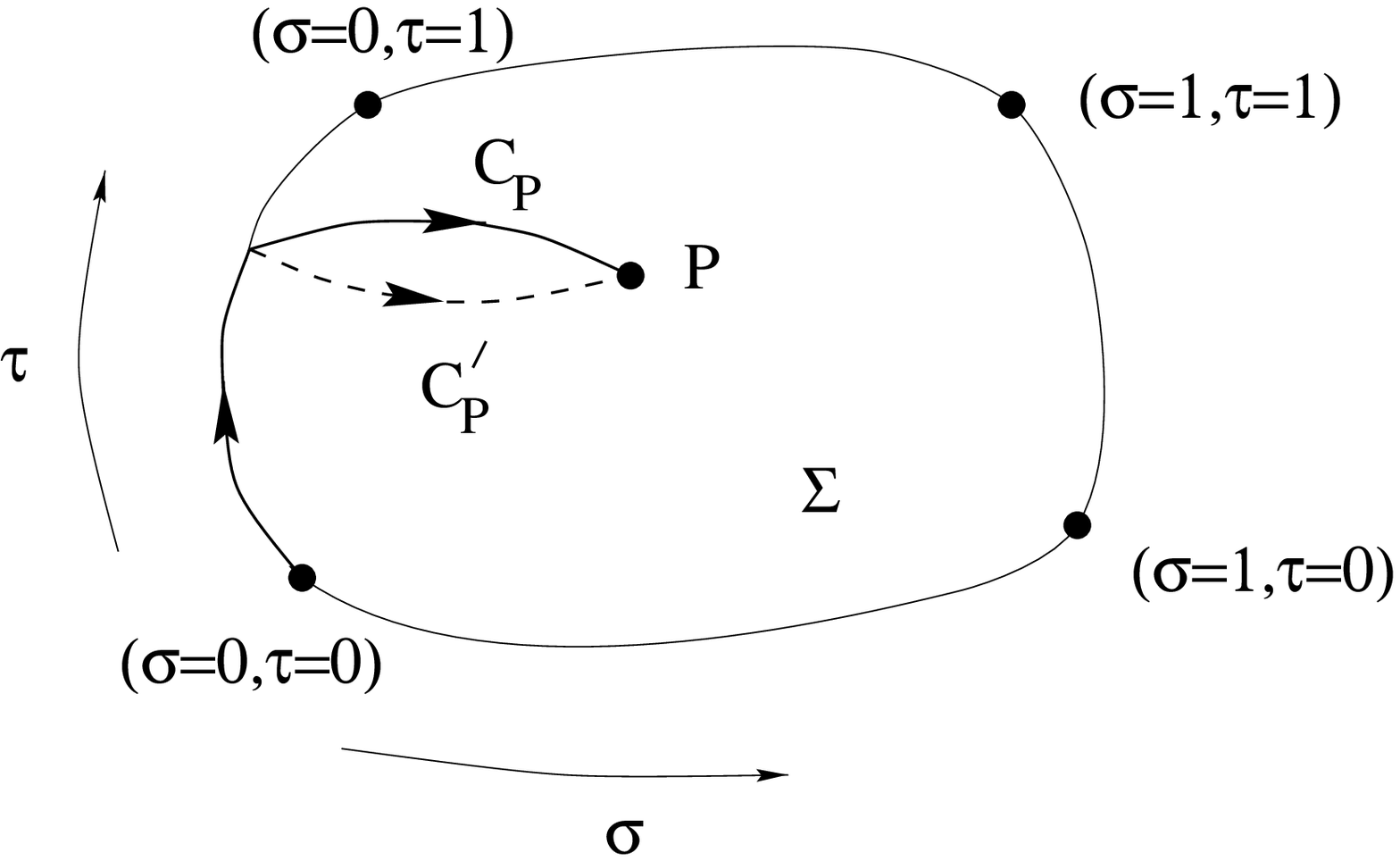}
\hspace{.5truecm}

\centerline{}
{\vspace{-8truemm}}
\caption{A parametrized surface $\Sigma$. The path $C_P$ consists of
two segments:  the first segment
$(\sigma =0={\rm const.}, \tau )$ is from $\tau=0$ to $\tau$ and 
the second segment $(\sigma , \tau = {\rm const.})$ is from  $\sigma = 0$
to $\sigma$.}
\label{ordering}
}

It can be checked that the gauge transformations (\ref{gauge1}--\ref{gauge3})
are compatible with the composition rule (\ref{funcom}) provided that
the composition rule for $R$ is  the same as that of $W$, namely
\be
R[\Sigma_2\circ \Sigma_1] = M[C]( R[\Sigma_2] ) 
M[C] V[\Sigma_2] M[C^{-1}] ( R[\Sigma_1] ).
\ee
More generally, consider a surface $\Sigma$ divided into $n$ smaller
surfaces $\Sigma_1,\ldots, \Sigma_n$. Let $C$ be the boundary of $\Sigma$.
Repeating the reasoning leading to \eq{funcom} we have
\be
M[C]= M[\C_1] i_{W[\Sigma_1]} V[\Sigma_1] M[\C_2] i_{W[\Sigma_2]} V[\Sigma_2]
M[\C_3] \cdots 
\ee
for some curves $\C_1, \C_2,\ldots$. From this equation we find
$$
W[\Sigma]= M[\C_1](W[\Sigma_1]) M[\C_1] V[\Sigma_1] M[\C_2](W[\Sigma_2]) 
\ldots ,
$$
\be
V[\Sigma] = M[\C_1] V[\Sigma_1] M[\C_2] V[\Sigma_2] M[\C_3] \ldots .
\label{lotcom}
\ee
It is easy to see that the gauge transformations (\ref{gauge1}--
\ref{gauge3}) are compatible with \eq{lotcom} provided that
$R[\Sigma]$ is composed out of $R[\Sigma_i]$ as follows:
\be
R[\Sigma]= M[\C_1](R[\Sigma_1]) M[\C_1] V[\Sigma_1] M[\C_2](R[\Sigma_2]) 
\ldots .
\ee
Thus $R$  should be composed by  the  rule of composition of $W$.

We now introduce new gauge transformations. These   
 are the transformations of $M$, $V$ and $W$ 
compatible with  \eq{fund}.

Let $\Lambda[P]$ be an $\au$-valued function of  point $P$. Let 
$C$ be a directed path from $P_1$ to $P_2$. The gauge transformation
of $M[C]$ reads
\be
{\tilde M}[C]= \Lambda[P_2] M[C] \Lambda[P_1]^{-1}.
\label{Mlambda}
\ee
When $P_1=P_2=P$ this equation becomes
\be
{\tilde M}[C]= \Lambda[P] M[C] \Lambda[P]^{-1}.
\ee
From this equation and  
\be
{\tilde M}[C] = i_{\tilde W} {\tilde V}
\ee
one finds 
\be
i_W V = \Lambda^{-1} i_{\tilde W} {\tilde V} \Lambda = i_{\Lambda^{-1}
({\tilde W})} \Lambda^{-1} {\tilde V} \Lambda .
\ee
Thus we propose the gauge transformations: 
$$
{\tilde V}[\Sigma,P]= \Lambda[P] V[\Sigma,P] \Lambda[P]^{-1},
$$
\be
{\tilde W}[\Sigma,P]= \Lambda[P](W[\Sigma,P]).
\label{glambda}
\ee

We now consider a new gauge transformation which is a finite generalization
of  the infinitesimal transformation considered in \cite{BM}. 
The transformation
reads 
\be
{\tilde M}[C] = i_{\Z[C]} M[C],
\label{Mtrans}
\ee
where $\Z[C]$ is a $G$-valued functional of $C$. The composition
rule for $\Z$ can be inferred from the following chain of equations:
\bea
i_{\Z[C_2\circ C_1]} M[C_2\circ C_1]&=& {\tilde M}[C_2\circ C_1]\non
&=& {\tilde M}[C_2] {\tilde M}[C_1] \non
&=& i_{\Z[C_2]} M[C_2] i_{\Z[C_1]} M[C_1] \non
&=& i_{\Z[C_2]} i_{M[C_2] (\Z[C_1])} M[C_2\circ C_1] .
\eea
This equation suggests the following
composition rule for  $\Z$:
\be
\Z[C_2 \circ C_1] = \Z[C_2] M[C_2] (\Z[C_1]).
\label{zcompose}
\ee
If a Lie($G$)-valued 1-form $\zeta$ is given, $\Z[C]$ for an open path $C$
can be constructed as follows. Let us divide $C$ into $n$ small subpaths
as in figure \ref{zcontour}(a). Applying  \eq{zcompose} we find
\bea
\Z[C] &=& \Z[C_n] \cdot M[C_n](Z[C_{n-1}]) \cdot M[C_n\circ C_{n-1}](\Z[C_{n-2}])
\cdots \non
&&~~M[C_n \circ C_{n-1} \cdots C_2] (\Z[C_1]) \non
&\approx& (1 + \zeta_{\mu}[P_{n}] dx^{\mu}) (1 + M[C_n](\zeta_{\mu}[P_{n-1}])dx^{\mu})
\cdots \non
&&~~(1+ M[C_n \circ C_{n-1} \cdots C_2]( \zeta_{\mu}[P_1] )dx^{\mu}).
\eea
In the large $n$ limit we thus find
\be
\Z[C] = {\hat P}~\ex \LB \int_C  M[C''] (\zeta_{\mu}[P]) dx^{\mu} \RB ,
\ee
where $C''$ and $P$ are as in figure \ref{zcontour}(b), and ${\hat P}$ is
the path ordering operator.

\FIGURE[ht]{
\includegraphics[width=10truecm]{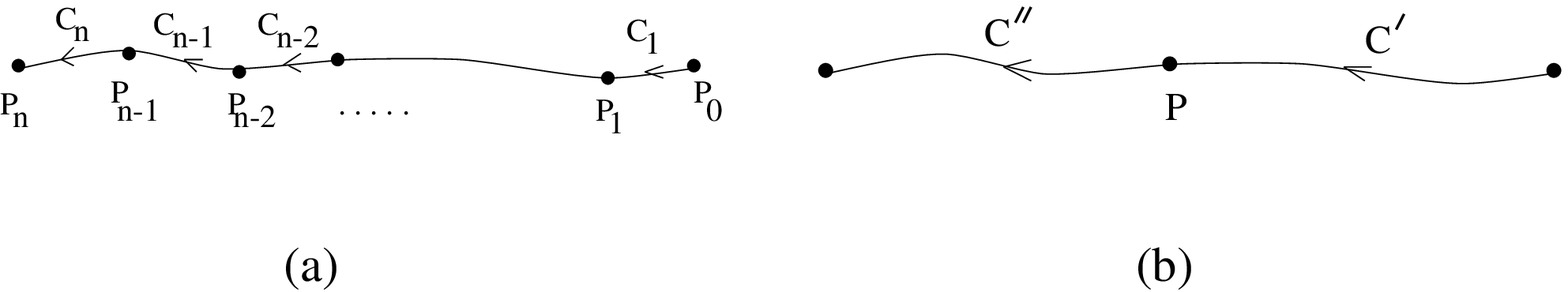}
\hspace{.5truecm}

\centerline{}
{\vspace{-8truemm}}
\caption{(a) The path $C$ is divided into $n$ small subpaths: 
$C= C_n \circ C_{n-1} \cdots \circ C_1$. (b) The point $P$ divides
$C= C'' \circ C'$.
}
\label{zcontour}
}

A choice of  transformation of $V$ and $W$ compatible with \eq{fund} 
and \eq{Mtrans} is
$$
{\tilde V}[\Sigma,C]= V[\Sigma,C],
$$
\be
{\tilde W}[\Sigma,C]= \Z[C] W[\Sigma,C].
\label{VWtrans}
\ee
Infinitesimal versions of these transformations 
agree with the transformations that can be derived from 
\cite{BM}. Let us consider an infinitesimal surface $\delta \Sigma$
with the area element
$\sigma^{\mu\nu}$. Assume that $M[C]\in \au$ is an inner automorphism
given by
\bea
M[C] (g) &=& {\hat P} \ex \LB \int_C \mu \RB  ~g~ {\hat P}\ex \LB -\int_C \mu
\RB \non
&=& {\hat P} \ex \LB \int_C \mu_{\scriptstyle{\rm adjoint}} \RB (g),~~~~~\forall g\in G,
\eea
where $\mu$ is a Lie($G$)-valued 1-form.
From \eq{VWtrans} and 
\be
W[\delta \Sigma] \approx 1 + B_{\mu\nu} \sigma^{\mu\nu},
\ee
one can find 
the transformation of the 2-form $B$: 
\be
{\tilde B} = B + d\zeta - {1\over 2} [\zeta , \zeta] - [\mu, \zeta] .
\ee
The transformation of $B$ corresponding to eqs.(\ref{gauge1},\ref{gauge2})
 reads
\be
{\tilde B}_{\scriptstyle{\rm ab}}= B_{\scriptstyle{\rm ab}},~~~~
{\tilde B}_{\scriptstyle{\rm nonab}} = B_{\scriptstyle{\rm nonab}} - \rho ,
\label{gaugeB}
\ee
where $\rho$ is a Lie($G$)-valued 2-form defined in
\be
R[\delta \Sigma] \approx 1 + \rho_{\mu\nu} \sigma^{\mu\nu} .
\ee
Eq.(\ref{gaugeB}) agrees with the transformations that can be derived
from \cite{BM}.

Unlike the gauge transformations (\ref{gauge1}--\ref{gauge3},
\ref{glambda}), the transformation 
(\ref{VWtrans}) is not compatible with the
composition rule (\ref{funcom}). To find the correct transformation,  
$\Z[C]$ in \eq{VWtrans} should  be
`smeared' over the surface $\Sigma$. We give an explicit 
formula for the  gauge transformation of $V[\Sigma]$. It reads
\be
{\tilde V}[\Sigma] = {\hat P}_{\tau} \exp \LB \int_{\Sigma} i_{\Z[C_P]} M[C_P] v[P] M[C_P^{-1}] i_{\Z[C_P]^{-1}} \RB .
\label{smeared}
\ee

\section{Comments}

\noindent
$\bullet$ We found three kinds of gauge transformations of $M$, $V$ and $W$.
These are $\Lambda[P]$-transformations (\ref{Mlambda},\ref{glambda}),
$R[\Sigma]$-transformations (\ref{gauge1}--\ref{gauge3}) and
$\Z [C]$-transformations (\ref{Mtrans},\ref{VWtrans}). Eq.(\ref{VWtrans})
is valid only for infinitesimal surfaces and should be replaced by
a `smeared' version \eq{smeared}.

\noindent
$\bullet$ The ambiguity in surface-ordering necessitates the introduction of
gauge transformations which compensate the ambiguity. Locally this amounts
to the transformation \eq{gaugeB}.
The number of gauge degrees of freedom present in a NWS is enormous.
Thus NWS may be relevant to a topological string theory
describing   topological sectors of the non-abelian string of \cite{six}.

\noindent
$\bullet$ Infinitesimal version of \eq{funcom} can be derived from
the composition rule for the natural transformation $K$ in figure 
\ref{natural}. 

\noindent
$\bullet$ We defined NWS on a local trivial patch. To define NWS
globally one should cover the manifold with an atlas $\{ U_{\alpha}\}$
and introduce $W_{\alpha}, V_{\alpha}, M_{\alpha}$ for each patch $U_{\alpha}$.
As usual the quantities on the overlaps $U_{\alpha \beta} = U_{\alpha}\cap 
U_{\beta}$ are related by the gauge transformations. An analysis
of global issues will be carried out elsewhere.

\noindent
$\bullet$ We defined NWS with  the disk topology. A generalization
to higher-genus surfaces will be discussed  elsewhere.

\vspace{5mm}

\noindent
{\it Note added} 
 
After submitting the original version of this paper
to hep-th, the work \cite{integra} was brought to our attention. 
In \cite{integra}  an equation similar to \eq{Vsurface} was taken as a
definition of Wilson surface. The case considered in \cite{integra}
corresponds, in our notation, to the $C$-independent $M[C]$.
The surface-ordering ambiguities are absent in this case. 
For a list of
miscellaneous   work on non-abelian 2-form theories, see \cite{miscel}.

\vspace{1cm}

\noindent
{\bf Acknowledgments}

This work was supported by the DOE under grant No. DE-FG03-92ER40701, by 
the Research Foundation under NSF grant
PHY-9722101 and by CRDF Award RP1-2108.

\end{document}